\documentclass[
amsmath,amssymb,a4paper,
prd,
superscriptaddress,
]{revtex4}
\usepackage{graphicx}
\usepackage[dvipdfm]{hyperref}

\begin{document}
\title{Numerical simulation of Quasi-Normal Modes \\
in time-dependent background}
\author{Li-Hui Xue}
\author{Zai-Xiong Shen}
\affiliation{Surface Physics Laboratory (National Key Lab), Fudan University, Shanghai 200433, People's Republic of China}
\author{Bin Wang}
\affiliation{Department of Physics, Fudan University, Shanghai 200433, People's Republic of China}
\author{Ru-Keng Su}
\affiliation{China Center of Advanced Science and Technology (World Lab), P. O. Box 8730, Beijing 100080, People's Republic of China}
\affiliation{Department of Physics, Fudan University, Shanghai 200433, People's Republic of China}
\date{\today}
\begin{abstract}
We study the massless scalar wave propagation in the time-dependent Schwarzschild black hole background. We find that the Kruskal coordinate is an appropriate framework to investigate the time-dependent spacetimes. A time-dependent scattering potential is derived by considering dynamical black hole with parameters changing with time. It is shown that in the quasi-normal ringing both the decay time-scale and oscillation are modified in the time-dependent background.
\end{abstract}
\maketitle

\section{Introduction
\label{sec:1}}
The study of the Quasi-Normal Modes (QNM) developing outside black holes plays an important role in black hole physics and astrophysics. In virtue of previous works, we now have the schematic picture regarding the dynamics of waves outside a black hole. A static observer outside the black hole can indicate three successive stages of the wave evolution. The first stage is the exact shape of the wave front, which depends on the initial pulse. This stage is followed by the quasi-normal ringing, which has characteristic frequencies directly connected to the parameters of the black hole. This quasi-normal ringing is regarded as a direct evidence of the existence of black holes and is expected to be detectable by future gravitational wave detectors. Finally, at late time, the quasi-normal ringing is replaced by a relaxation process, which is known as the wave tails. Wave tail is of interest in its own right since it has various applications from the no-hair theorem to the mass inflation scenario.

Quasi-normal modes of massless scalar, electromagnetic, and gravitational fields in asymptotically flat spacetimes have been studied for over thirty years. For a review, see \cite{1}. It was shown that the quasi-normal ringing decays exponentially with time. The oscillation period and the decay time scale are associated with the real part and the imaginary part respectively of the so called complex quasi-normal frequency \cite{2}. The tail phenomenon was first investigated by Price \cite{3} for both Schwarzschild and Reissner-Nordstr\"{o}m black hole. It was shown that for an observer at a fixed position the field dies off with a power-law tail. This behavior was confirmed by numerical studies \cite{4}. The evolution of massive scalar field in the asymptotically flat black hole background is also of interest. Recently this topic has attracted a lot of attention \cite{5}.

The study of wave dynamics is not limited in the asymptotically flat spacetimes. Motivated by the inflation, the quasi-normal modes in de Sitter (dS) space were studied \cite{6,7}. One of the most significant differences between the QNMs in the dS space and the QNMs in the asymptotically flat space is that in the dS space the oscillation always decays exponentially, the power-law tail is not in presence. Inspired by the discovery of the AdS/CFT correspondence, the investigation of the quasi-normal modes in the Anti de Sitter (AdS) black hole has been appealing \cite{8,9,10,11} recently. It was shown that the decay of the perturbation is always exponential for the AdS black holes.

Another intriguing aspect of studying QNM has been suggested recently. It was argued that QNM could be a probe to investigate the quantum properties of black holes. The first speculation was put forward by Hod who pointed out that the QNMs of a Schwarzschild black hole conform to the existence of the Bekenstein-Hawking temperature and imply emission in terms of a fixed area quantum \cite{12}. This study was subsequently extended to other black holes such as the d-dimensional spherically symmetric black holes \cite{13}, near extreme Schwarzschild-de Sitter and Kerr black holes \cite{14}, etc \cite{15}. Though this kind of consideration is still somewhat speculative, it possibly relates classical, macroscopic properties to quantum information of black holes.

All these previous works on quasi-normal modes have so far been restricted to time-independent black hole backgrounds. It should be realized that, for a realistic model, the black hole parameters change with time. A Schwarzschild black hole gaining or losing mass via absorption or evaporation is a good example. The more intriguing investigation of the black hole QNM calls for a systematic analysis of time-dependent spacetimes. Recently the late time tails under the influence of a time-dependent scattering potential has been explored in \cite{16}, where the tail structure was found modified due to the temporal dependence of the potential. The motivation of the present paper is to explore the modification to the QNM in the time-dependent spacetimes. Instead of putting an effective time-dependent scattering potential by hand as done in \cite{16}, we will introduce the time-dependent potential in a natural way by considering dynamic black holes with black hole parameters changing with time due to absorption and evaporation processes. We will study numerically the temporal evolution of massless scalar field perturbation, especially the QNM in different time-dependent situations.

The outline of this paper is as follows. In Sec. \ref{sec:2} we first go over the conventional treatment for numerical study of the wave propagation in the stationary black hole background. We will point out the difficulty if it is to be applied to the time-dependent cases. Then we will introduce a new treatment by employing the Kruskal coordinates which we find to be advantageous. In Sec. \ref{sec:3}, we present the numerical study of the QNM in the time-dependent background. Sec. \ref{sec:4} is a brief summary and discussion.

\section{Field evolution in time-independent background
\label{sec:2}}
\subsection{The conventional treatment}
In this section, we first present a brief review on the conventional treatment of the evolution of massless scalar field propagating in a stationary Schwarzschild black hole background. For more details, please refer to \cite{4}.

The metric of a stationary (time-independent) Schwarzschild black hole in the general coordinates is given by
\begin{equation}
ds^2 = -f dt^2 + f^{-1}dr^2 +r^2 d \Omega^2,
\label{eq:1}
\end{equation}
where $f = 1 - 2M/r$. The propagation of waves in curved spacetimes is governed by the Klein-Gordon equation
\begin{equation}
\Box \Phi = 0.
\label{eq:2}
\end{equation}
Substituting the metric, Eq. (\ref{eq:2}) becomes
\begin{equation}
\frac{r^2}{f}\Phi_{,tt} - \left( r^2 f_{,r} + 2rf \right)\Phi_{,r} - r^2 f \Phi_{,rr} - \cot \theta \Phi_{,\theta} - \Phi_{,\theta \theta} - \frac{1}{\sin^2 \theta} \Phi_{,\phi \phi} = 0.
\label{eq:3}
\end{equation}

Since the background is spherically symmetric, each multipole of a perturbation field evolves separately. Hence one can write
\begin{equation}
\Phi(t,r,\theta,\phi) = \psi_{l}(t,r) Y_{lm}(\theta,\phi)/r,
\label{eq:4}
\end{equation}
and obtains a wave equation for each multipole moment
\begin{equation}
f^{-1}\psi_{,tt} - f \psi_{,rr} - f_{,r} \psi_{,r} + \left[ \frac{l(l+1)}{r^2} + \frac{f_{,r}}{r} \right]\psi = 0.
\label{eq:5}
\end{equation}
Using the tortoise coordinate $r^*$ defined by
\begin{equation}
dr^* = f^{-1} dr,
\label{eq:6}
\end{equation}
the wave equation becomes
\begin{equation}
\psi_{,tt} - \psi_{,r^* r^*} + V(r)\psi = 0,
\label{eq:7}
\end{equation}
where the effective potential
\begin{equation}
V(r) = f \left[ \frac{l(l+1)}{r^2} + \frac{f_{,r}}{r} \right].
\label{eq:8}
\end{equation}

Eq. (\ref{eq:7}) can be recast as 
\begin{equation}
\psi_{,uv} + \frac{1}{4}V(r)\psi = 0,
\label{eq:9}
\end{equation}
where the null coordinates 
\begin{equation}
u = t - r^*, \; v = t + r^*.
\label{eq:10}
\end{equation}
To write Eq. (\ref{eq:9}) into the discrete form, one can construct a functional
\begin{equation}
I[\psi] = \iint dudv F[u,v,\psi,\psi_{,u},\psi_{,v}],
\label{eq:11}
\end{equation}
and the differential equation Eq. (\ref{eq:9}) is recovered from the functional variation
\begin{equation}
\frac{\delta I[\psi]}{\delta \psi} = 0,
\label{eq:12}
\end{equation}
or
\begin{equation}
\frac{\partial}{\partial u}\left( \frac{\partial F}{\partial \psi_{,u}} \right) + \frac{\partial}{\partial v}\left( \frac{\partial F}{\partial \psi_{,v}} \right) - \frac{\partial F}{\partial \psi} = 0,
\label{eq:13}
\end{equation}
if the function $F$ takes the form
\begin{equation}
F[u,v,\psi,\psi_{,u},\psi_{,v}] = \psi_{,u} \psi_{,v} - \frac{1}{4}V(r)\psi^2.
\label{eq:14}
\end{equation}

Using the three point formula for the first-order derivatives and the trapezoid rule, one can convert Eq. (\ref{eq:11}) into a summation form
\begin{equation}
I[\psi] = \Delta u \Delta v \sum_{i,j}\left[ \frac{(\psi_{i+1,j} - \psi_{i-1,j}) (\psi_{i,j+1} - \psi_{i,j-1})}{\Delta u \Delta v} - \frac{V(r)\psi_{i,j}^2}{4} \right],
\label{eq:15}
\end{equation}
where $\psi_{i,j}$ denotes $\psi(u_i,v_j)$.

Now each discrete $\psi_{i,j}$ is treated as an independent variable, and the functional variation $\delta I[\psi] / \delta \psi = 0$ becomes a partial derivative
\begin{equation}
\frac{\partial I[\psi_{i,j}]}{\partial \psi_{i,j}} = 0.
\label{eq:16}
\end{equation}
Substituting Eq. (\ref{eq:15}), Eq. (\ref{eq:16}) becomes
\begin{equation}
\psi_{i-1,j+1} + \psi_{i+1,j-1} - \psi_{i-1,j-1} - \psi_{i+1,j+1} - \frac{1}{4} \Delta u \Delta v V(r) \psi_{i,j} = 0.
\label{eq:17}
\end{equation}
Taking the following notations
\begin{equation}
\begin{split}
\psi_N = \psi_{i+1,j+1}, \; &\psi_S = \psi_{i-1,j-1}, \\
\psi_E = \psi_{i+1,j-1}, \; &\psi_W = \psi_{i-1,j+1},
\end{split}
\label{eq:18}
\end{equation}
and replacing the midway quantity between two neighboring points by the average of two lattice points
\begin{equation}
\psi_{i,j} = \frac{\psi_W + \psi_E}{2},
\label{eq:19}
\end{equation}
one arrives at
\begin{equation}
\psi_N = \psi_W + \psi_E - \psi_S - \Delta u \Delta v V(r) \frac{\psi_W + \psi_E}{8} = 0.
\label{eq:20}
\end{equation}

It is straightforward to setup a rectangular grid of $u$ and $v$. Starting from initial data on $u = u_0$ and $v = v_0$, integration can proceed to the northeast, as described in \cite{4} and numerical result of Eq.(\ref{eq:9}) can be obtained.

The conventional treatment is powerful in studying the wave propagation in stationary black hole backgrounds, however it is difficult to be extended to investigate the time-dependent spacetimes. For a dynamic Schwarzschild black hole, the mass of the black hole is a function of time $M = M(t)$, then we get another term $-(f_{,t}/f^2)\psi_{,t}$ in Eq. (\ref{eq:5}) for each multipole moment. Due to the appearance of the first-order derivative associated with time, the neat expression Eq. (\ref{eq:7}) and Eq. (\ref{eq:9}) are impossible to be obtained. Such an equation is difficult to handle. Besides the tortoise coordinate $r^*$ now is not only a function of $r$, but also a function of $t$. These difficulties  make the conventional numerical treatment no longer appropriate in the time dependent background.

\subsection{Wave equation in the Kruskal coordinates}
To avoid these difficulties, we use the Kruskal coordinates instead of the general coordinates Eq. (\ref{eq:1}). As we know, the metric terms of the time-like coordinate $\tau$ and the space-like coordinate $\rho$ are the same in the Kruskal coordinates, hence the mass of the black hole $M(t)$ only appears in the angular part of the wave equation. Thus in the time-dependent case, the investigation will be easier. This will be exhibited in Sec. \ref{sec:3}. Here we first present the investigation in the stationary black hole background in order to test the validity of applying the Kruskal coordinates in studying the wave propagation.

In the Kruskal coordinates, the metric for the Schwarzschild black hole is given by
\begin{equation}
ds^2 = \frac{32M^3}{r}e^{-r/2M}(d \tau^2-d \rho^2)-r^2 d \Omega^2,
\label{eq:21}
\end{equation}
where $\tau$ and $\rho$ are the time-like and space-like coordinates respectively. In the region $r \ge 2M$, they are defined by
\begin{equation}
\tau = \sqrt{\frac{r}{2M}-1} \, e^{r/4M}\sinh \frac{t}{4M}, \; \rho = \sqrt{\frac{r}{2M}-1} \, e^{r/4M}\cosh \frac{t}{4M}.
\label{eq:22}
\end{equation}
The propagation of the massless scalar field is governed by the wave equation
\begin{multline}
2r \sin \theta r_{,\tau} \Phi_{,\tau} + r^2 \sin \theta \Phi_{,\tau \tau} - 2r \sin \theta r_{,\rho} \Phi_{,\rho} - r^2 \sin \theta \Phi_{,\rho \rho} \\
- \frac{32M^3}{r}\cos \theta e^{-r/2M} \Phi_{,\theta} - \frac{32M^3}{r} \sin \theta e^{-r/2M} \Phi_{,\theta \theta} - \frac{32M^3}{r \sin \theta} e^{-r/2M} \Phi_{,\phi \phi} = 0.
\label{eq:23}
\end{multline}
Since the background is spherically symmetric, we can write $\Phi(\tau,\rho,\theta,\phi) = \psi_{l}(\tau,\rho) Y_{lm}(\theta,\phi)/r$, and we get the wave equation for each multipole moment
\begin{equation}
\psi_{,\tau \tau} - \psi_{,\rho \rho} + \left[ \frac{32M^3 l(l+1) e^{-r/2M}}{r^3} - \frac{r_{,\tau \tau}}{r} + \frac{r_{,\rho \rho}}{r} \right] \psi = 0.
\label{eq:24}
\end{equation}

Analogous to the null coordinates used in \cite{4}, we make the following variable transformations
\begin{equation}
u = \tau -\rho, \; v = \tau + \rho,
\label{eq:25}
\end{equation}
and the wave equation becomes
\begin{equation}
\psi_{,uv} + V \psi = 0,
\label{eq:26}
\end{equation}
where
\begin{equation}
V = \frac{8M^3 l(l+1) e^{-r/2M}}{r^3} - \frac{r_{,uv}}{r}.
\label{eq:27}
\end{equation}
The expression of $r_{,uv}$ is simple when $M$ is a constant. For the stationary black hole, we have
\begin{equation}
r_{,uv} = -\frac{16M^4 e^{-r/2M}}{r^3}.
\label{eq:28}
\end{equation}

It is straightforward to write Eq. (\ref{eq:26}) into the discrete form
\begin{equation}
\psi_N = \psi_W + \psi_E - \psi_S -\Delta u \Delta v V \frac{\psi_W + \psi_E}{2}.
\label{eq:29}
\end{equation}
The only difficulty of developing the code is to retrieve $r$ and $t$ from Eq. (\ref{eq:22}) and Eq. (\ref{eq:25}) for each point on the grid. In a typical setup, the approximate range of $u$ and $v$ are from $-10^0$ to $-10^{-50}$ and from $10^1$ to $10^{50}$ respectively. In order to finish the integration over such a big region in a reasonable duration of time while sticking to an acceptable precision, we divide the whole grid into small blocks. The step length $\Delta u$ and $\Delta v$ vary from block to block in accordance with the requirement. Starting from the southwest, the small blocks are resolved one by one to the northeast, the output of each resolved block becomes the input of the forthcoming blocks. We use $\psi(v = v_0) = 0$ and a Gaussian pulse on $u = u_0$ as the initial data for all of our calculations.

\begin{figure}
\includegraphics{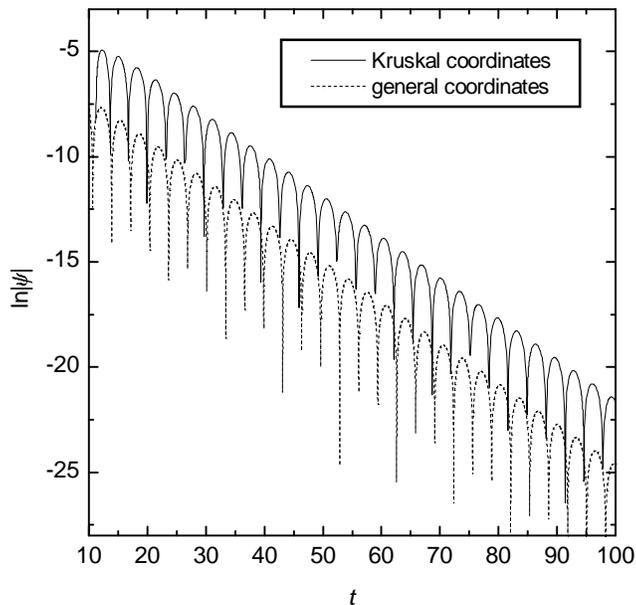}
\caption{Temporal evolution of the field $\psi$ at a fixed radius $r = 2$. The mass of the black hole $M = 0.5$ and the multipole index $l = 2$. The upper curve is obtained from calculation done in the Kruskal coordinates, the lower curve is calculated in the general coordinates.
\label{fig:1}}
\end{figure}

The numerical evolution of the $l = 2$ mode is displayed in Fig. \ref{fig:1}. For comparison, we also exhibit the curve obtained in the general coordinates using the conventional treatment. Since the amplitude of the oscillations is physically irrelevant, we concern ourselves on the quasi-normal frequencies of the oscillations. The real part of the quasi-normal frequency $\omega_R$ and the imaginary part of the quasi-normal frequency $\omega_I$ read from Fig.1 and the quasi-normal frequencies of a few other modes are listed in Table \ref{tab:1}. As one can see from the table, numerical result got by employing the Kruskal coordinates agrees well to that of the conventional treatment. This shows that our treatment using the Kruskal coordinate is valid to investigate the wave propagation in the black hole background.

\begin{table}
\caption{Quasi-normal frequencies for the massless scalar perturbation in the stationary Schwarzschild background. The mass of the black hole $M = 0.5$.
\label{tab:1}}
\begin{ruledtabular}
\begin{tabular}{ccccc}
&\multicolumn{2}{c}{\textbf{General Coordinates}}&\multicolumn{2}{c}{\textbf{Kruskal Coordinates}} \\
$l$&$\omega_R$&$\omega_I$&$\omega_R$&$\omega_I$ \\
1&0.586\footnote{It is $0.56$ in \cite{4}.}&-0.1956\footnote{It is $-0.19$ in \cite{4}.}&0.587&-0.1947 \\
2&0.968&-0.1932&0.968&-0.1932 \\
3&1.351&-0.1926&1.355&-0.1926 \\
4&1.736&-0.1924&1.735&-0.1922
\end{tabular}
\end{ruledtabular}
\end{table}

\section{Field evolution in time-dependent backgrounds
\label{sec:3}}
If the mass of the black hole has a general dependence on $t$, then $M$ is a function of $\tau$ and $\rho$ in the Kruskal coordinate. Since the metric $g_{\tau\tau}=g_{\rho\rho}$, the mass $M$ only appears in the angular part of the wave equation. This is a big advantage of using the Kruskal coordinates in study of the time-dependent case. The wave equation has the same form as 
Eq. (\ref{eq:26}), however now the second-order derivative $r_{,uv}$ is replaced by
\begin{equation}
r_{,uv} = -\frac{16M^4 e^{-r/2M}}{r^3} + \frac{rM_{,uv}}{M} - \frac{4M(M_{,u}u - M_{,v}v) e^{-r/2M}}{r}.
\label{eq:30}
\end{equation}
The derivatives $M_{,u}$, $M_{,v}$, and $M_{,uv}$ are given by
\begin{subequations}
\begin{align}
&M_{,u} = \frac{\dot{M}}{2}(t_{,\tau} - t_{,\rho}), \\
&M_{,v} = \frac{\dot{M}}{2}(t_{,\tau} + t_{,\rho}), \\
&M_{,uv} = \frac{\ddot{M}}{4}(t_{,\tau}^2 - t_{,\rho}^2) + \frac{\dot{M}}{4}(t_{,\tau \tau} - t_{,\rho \rho}),
\end{align}
\label{eq:31}
\end{subequations}
where $\dot{M}$ and $\ddot{M}$ are the first-order and second-order derivatives with respect to $t$. The terms $t_{,\tau}$, $t_{,\rho}$, $t_{,\tau \tau}$, and $t_{,\rho \rho}$ can be obtained directly from  Eq. (\ref{eq:22})
\begin{subequations}
\begin{align}
&t_{,\tau} = \frac{4M^2 \cosh^2\frac{t}{4M}}{\rho (M - \dot{M}t)}, \\
&t_{,\rho} = -\frac{4M^2 \sinh^2\frac{t}{4M}}{\tau (M - \dot{M}t)}, \\
&t_{,\tau \tau} = \frac{t_{,\tau}^2 \left[ \tanh\frac{t}{4M}(M - \dot{M}t)^2 + 2M^2\ddot{M}t + 4M \dot{M}(M - \dot{M}t) \right]}{2M^2(M - \dot{M}t)}, \\
&t_{,\rho \rho} = \frac{t_{,\rho}^2 \left[ \coth\frac{t}{4M}(M - \dot{M}t)^2 + 2M^2\ddot{M}t + 4M \dot{M}(M - \dot{M}t) \right]}{2M^2(M - \dot{M}t)}.
\end{align}
\label{eq:32}
\end{subequations}

\begin{figure}
\includegraphics{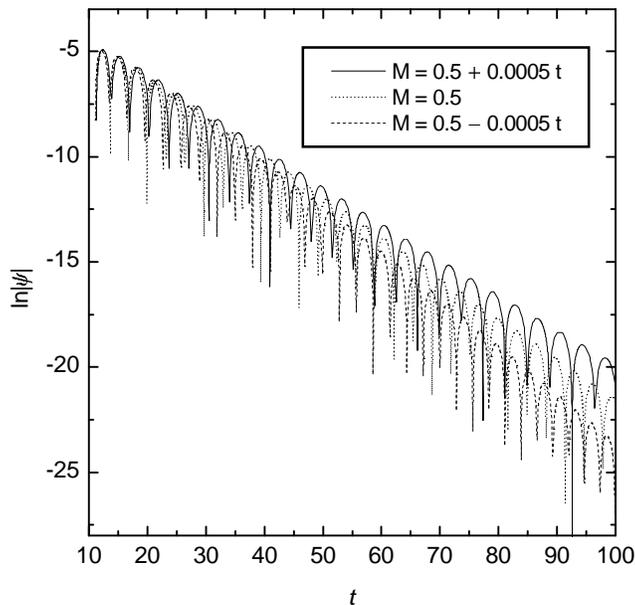}
\caption{Temporal evolution of the field in the background of a Schwarzschild black hole for $l = 2$ and $r = 2$. The mass of the black hole $M(t) = M_0 \pm at$, where $M_0 = 0.5$ and $a = 5 \times 10^{-4}$ are constant coefficients. The field evolution for $M = M_0 + at$ and $M = M_0 - at$ are shown as the top curve and the bottom curve respectively. For comparison, the oscillations for $M = M_0$ are also displayed (the middle curve).
\label{fig:2}}
\end{figure}

\begin{figure}
\includegraphics{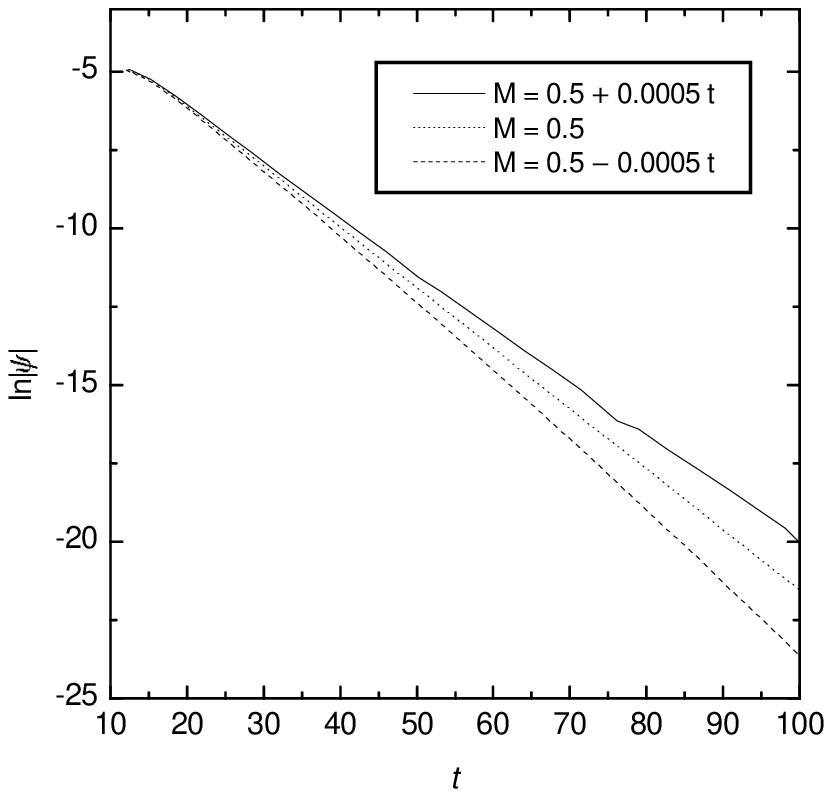}
\caption{The same plot as Fig. \ref{fig:2}. For clarity, the connected maxima of the oscillations are displayed. The imaginary frequency $\omega_I$ can be read from the decay time scale which is proportional to $1/\omega_I$.
\label{fig:3}}
\end{figure}

\begin{figure}
\includegraphics{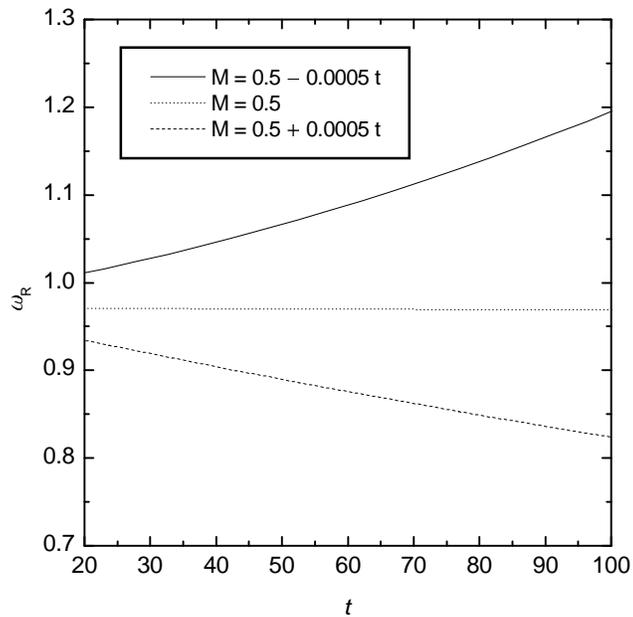}
\caption{The frequency $\omega_R$ determined from the oscillation period in Fig. \ref{fig:2} is shown as a function of $t$. 
\label{fig:4}}
\end{figure}

We now present the result of our numerical calculations in the time-dependent black hole background. In the first series of simulations, we consider the simple situation by choosing the mass of the black hole $M = M_0 \pm at$, where $M_0$ and $a$ are constant coefficients. The results are shown in Fig. \ref{fig:2}, \ref{fig:3}, and \ref{fig:4}. The modification to the QNM due to the time-dependent background is clear. When $M$ increases linearly with respect to $t$, the decay becomes slower compared to the stationary case, which corresponds to say that the $\omega_I$ is not a constant and decreases with respect to $t$ in the time-dependent situation. The oscillation period is no longer a constant as that for the stationary black hole, it becomes longer with the increase of time. In other words, the real part of the quasi-normal frequency $\omega_R$ decreases with the increase of time. When $M$ decreases linearly with respect to $t$, compared to the situation in the stationary black hole, we have observed that the decay becomes faster and the oscillation period becomes shorter which corresponds to both $\omega_I$ and $\omega_R$ increase with the increase of time.

\begin{figure}
\includegraphics{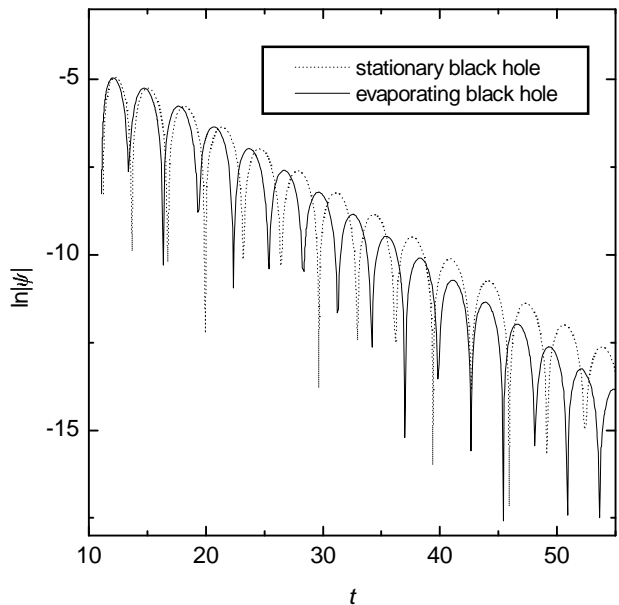}
\caption{Temporal evolution of the field $\psi$ in the background of a evaporating Schwarzschild black hole for $l = 2$ and $r = 2$ (lower curve). We use $\alpha_0 = 2.011 \times 10^{-4}$ \cite{17} and $b = 207.2$ in our calculation. For comparison, the $l = 2$ mode in the stationary background with $M = 0.5$ is also displayed (upper curve).
\label{fig:5}}
\end{figure}

\begin{figure}
\includegraphics{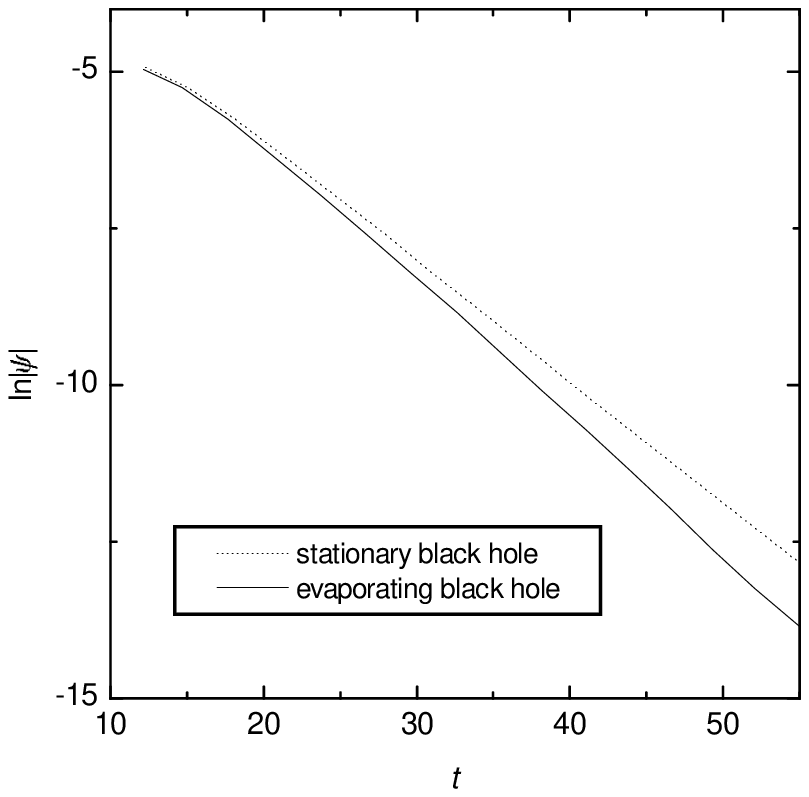}
\caption{Same as Fig. \ref{fig:5}. Only the connected maxima of the oscillations are displayed. The imaginary frequency $\omega_I$ can be read from the decay time scale which is proportional to $1/\omega_I$. 
\label{fig:6}}
\end{figure}

\begin{figure}
\includegraphics{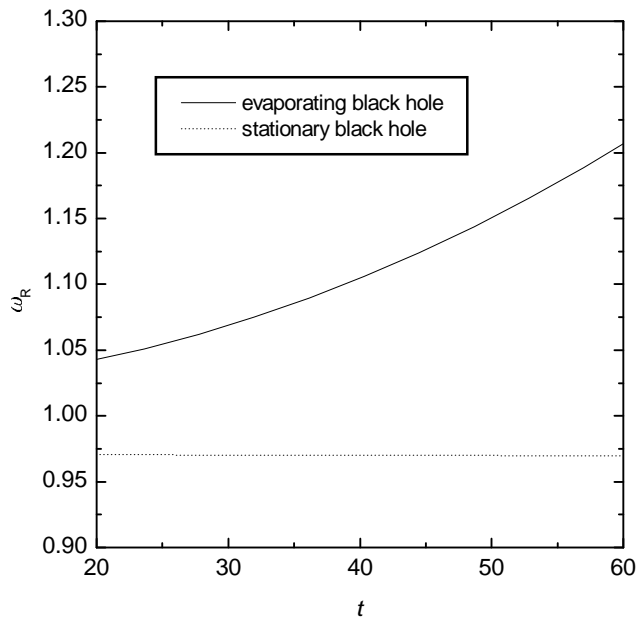}
\caption{The frequency $\omega_R$ determined from the oscillation period in Fig. \ref{fig:5} is shown as a function of $t$ for the evaporating black hole. For the stationary black hole $\omega_R$ is a constant. 
\label{fig:7}}
\end{figure}

We have also extended our discussion to a more realistic model, an evaporating Schwarzschild black hole with the mass determined by  \cite{17}
\begin{equation}
\dot{M} = -\frac{\alpha_0}{M^2},
\label{eq:33}
\end{equation}
where $\alpha_0$ is a constant coefficient that takes different values for black holes with mass $M \gg 10^{17} \text{g}$ and those with mass $M \ll 10^{17} \text{g}$. Since the wave equation Eq. (\ref{eq:23}) is invariant if one makes the rescale $M \rightarrow aM$, $r \rightarrow ar$, and $t \rightarrow at$, the value of $\alpha_0$ is not very important here. One can obtain the mass of the black hole as a function of $t$ from Eq. (\ref{eq:33}) as
\begin{equation}
M(t) = \left[ 3 \alpha_0 (b - t) \right]^{1/3},
\label{eq:34}
\end{equation}
where $b$ is an arbitrary constant. Results of the numerical calculations are shown in Fig. \ref{fig:5}, \ref{fig:6}, and \ref{fig:7}. Different from the stationary black hole case, for the evaporating black hole both real and imaginary parts of  quasi-normal frequencies $\omega_I$ and $\omega_R$ increase with respect to $t$, in consistent with the simple case above.

\section{Summary and discussion
\label{sec:4}}
We have studied the evolution of the massless scalar field in the time-dependent Schwarzschild black hole background. We have found that the Kruskal coordinates is an appropriate framework to investigate the wave propagation in the time-dependent spacetimes. Despite imposing the effective time-dependent scattering potential in the wave equation by hand, in our study we have tried to derive the time-dependent potential in a natural way by considering dynamic black holes with black hole parameters changing with time. In our numerical study, we have found the modification to the QNM due to the temporal dependence of the black hole spacetimes. The decay and oscillation timescale are no longer constants with the evolution of time as that in the stationary black hole case. In the absorption process, when the black hole mass becomes bigger, both the real and imaginary parts of the quasi-normal frequencies decrease with the increase of $t$. However, in the evaporating process, when the black hole loses mass, both the real and imaginary parts of the quasi-normal frequencies increase with the increase of $t$.

\begin{figure}
\includegraphics{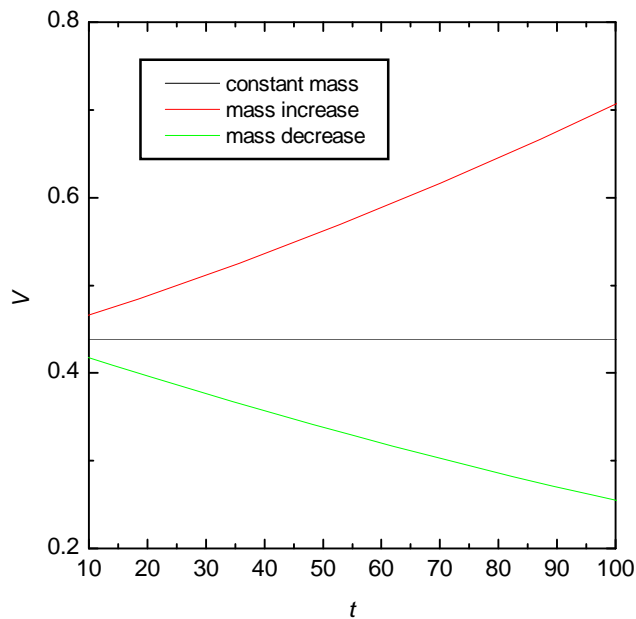}
\caption{The change of the effective potential with time for stationary black hole and dynamical black holes. 
\label{fig:8}}
\end{figure}

For the black hole evaporating case, black hole mass decreases with the increase of time and so does the effective potential, see Fig. \ref{fig:8}. The correspondence between the decrease of decay time scale and decrease of the effective potential with the evolution of time is qualitatively in consistent with the tail behavior exhibited in \cite{16}, where $V(x,t)=1/(x^{\alpha}t^{\beta})$ and the power-law indices found to be $-4.04$ and $-5.08$ for $\beta=1$ and $\beta=2$, respectively, showing that smaller potential (with bigger $\beta$ for the same $t$ and fixed $x$) leading to faster decay (bigger absolute value of power-law index).

\begin{acknowledgments}
This work was  partially supported by National Natural Science Foundation of China under  grant 10005004, 19947001, 10047005, 10235030 and the foundation of Ministry of Education of China.
\end{acknowledgments}

\end{document}